\documentclass[twocolumn]{article}

\usepackage{amssymb,amsmath}

\usepackage{graphicx}
\usepackage{hyperref}

\usepackage{newtxmath}

\usepackage[round,authoryear]{natbib}
\begin{document}

\title{Simulation of cloud drop collisions in ABC~flow}

\author{Torsten Auerswald\thanks{},
 \and Maarten H. P. Ambaum\\\\Dept. of Meteorology, University of Reading\\ Whiteknights campus Earley Gate PO Box 243, Reading RG6 6BB, UK.
 }

\date{}

\twocolumn[
  \begin{@twocolumnfalse}
    \maketitle
    \begin{abstract}
\noindent
Simulating the collision behaviour of cloud drops in a turbulent environment is numerically expensive. Because of the typical sizes of cloud drops, their motion is predominantly influenced by the smallest turbulent scales in the flow. We can exploit this property by using an Arnold--Beltrami--Childress (ABC) flow instead of a full direct numerical simulation (DNS) to simulate the turbulent effect on cloud drop collisions. This allows simulation of drop motion using much less computational resources than needed by DNS and therefore, allows for simulations with many more drops and larger model domains or more complex cloud physics. This is useful in cases where a specific process needs to be studied and the complexity and details of a realistic turbulent flow are of secondary interest.\\
It is shown that the simulations using ABC flow can faithfully reproduce collision statistics from much more comprehensive DNS calculations when using a similar Taylor microscale and root mean square velocity in the ABC and DNS flows. The results from the ABC flow are robust to changes in these two parameters within a reasonable range, even though adjustments to the parameters can be used to further tune the results.\\   

    \end{abstract}

  \end{@twocolumnfalse}
  ]
{
  \renewcommand{\thefootnote}%
    {\fnsymbol{footnote}}
  \footnotetext[1]{t.auerswald@reading.ac.uk}
}

\section{Introduction}

In cloud physics the collision kernel provides a measure for the ability of cloud drops to collide with each other. It is a crucial parameter for the understanding of many cloud physical processes. For example, the collision kernel is used in numerical weather forecast models to parameterise the development of the drop size spectrum \citep[e.g.][]{khain:15}, and in cloud physics to investigate the growth of cloud drops across the so called ``size-gap'' part of the drop size spectrum \citep{grabowski:13}. In this part of the spectrum drops are too large to benefit from condensational growth and too small to grow by gravitational collisions with other drops. In order to be able to produce rain drops this gap has to be bridged to allow drops to grow and form rain drops. It was shown in previous studies that the turbulent flow in a cloud can contribute significantly to enhancing the number of collisions for drops in this size range and therefore help drops overcome this size gap \citep{wang:09}.

In recent years, thanks to the increase in computational power, many studies have been conducted using Direct Numerical Simulation (DNS) to investigate the processes governing the collisions of cloud drops in a turbulent environment. These studies have helped e.g. understanding how turbulence affects the collision efficiency of drops depending on the dissipation rate \citep{wang:08}, and the different effects of dissipation rate and Reynolds number \citep{chen:16}. \citet{wang:05} studied the influence of turbulence on the aerodynamic interaction between drops; \citet{onishi:09} studied the influence of gravity on the collision kernel of drops in a turbulence field.

The above examples highlight the importance of DNS as a technique to study the interaction of cloud drops with turbulence. However, since DNS does not parameterise the smallest turbulence scales but resolves them explicitly, it has a large demand for computational resources. In this paper an alternative approach is proposed which is intended to help save computational resources and therefore free up computation time for simulating other cloud physical processes. Instead of simulating an unsteady turbulence field solving the Navier-Stokes equations with DNS, we demonstrate in this paper that reliable collision statistics can be reconstructed using a particular idealised frozen turbulence, instead of a fully resolved evolving turbulence field.

For the frozen turbulence, we use an ABC flow. This type of flow is stationary and consisting of several vortex tubes defined by trigonometric functions which are a stationary solution of the full Euler equations \citep{dombre:86}. Despite being laminar and steady, ABC flows can generate chaotic tracer trajectories \citep{wang:91} and therefore could be adequate for the simulation of collisions of cloud drops in a turbulent environment. This approach is much simpler and cheaper to compute, although by construction it excludes certain properties of atmospheric turbulence like unsteadiness or the turbulence spectrum. In this work the ability of the ABC flow field to simulate the behaviour of cloud drops falling under gravity in a turbulent environment is studied.

In Sec.~\ref{sec:method} the cloud model including the ABC flow, behaviour of drops and model setup is explained in detail. In Sec.~\ref{sec:results} results of ABC flow simulations are presented and analysed regarding their sensitivity to the ABC flow parameters length scale and velocity scale. These results are compared to the results from DNS simulations presented by \citet{ayala:08}. The ABC simulations provide collision kernels for 6 different drop size classes between 10 and 60~$\muup$m. Further simulations with 50 size classes in the same size interval have been conducted using the ABC flow. Those results are presented in Sec.~\ref{sec:full} and compared to the simulations with 6 size classes and the DNS by \citet{ayala:08}. In Sec.~\ref{sec:conc} the results are summarised and conclusions are presented as well as an outlook for future studies.

\section{Method}  
\label{sec:method}

In order to simulate the behaviour of cloud drops, the momentum equation of each drop is solved assuming spherical drops. This is a good assumption for the range of drop sizes we consider.

\subsection{Drop motion}
The drag on a drop was calculated using Stokes' law which states that for small Reynolds numbers the drag force experienced by a drop is proportional to the velocity difference between the drop and the air flow:
\begin{equation}
\vec{F}=6\pi\rho_{\mathrm{a}} \nu r\,\Delta \vec{v}.
\label{eq:fd}
\end{equation}
where $\rho_{\mathrm{a}}$ is the density of air, $r$ the drop radius, $\nu$ the kinematic viscosity of air and $\Delta \vec{v}$ the velocity difference between the drop and the air.

Terminal velocities resulting from Stokes' law are found to be in good agreement with measurements for small drops.

Comparing terminal velocities of falling drops in still air for different drop radii as measured by \citet{beard:69} and calculated from Eq.~\ref{eq:fd}, we find that for drop radii up to around 30~$\muup$m the Stokes' law gives an overestimate of the terminal velocity of up to 5\%, while for the largest simulated drop size of 60~$\muup$m, this overestimate increases to 25\%. The Reynolds number, based on drop radius, at terminal velocity for the  60~$\muup$m drop is 1.36.

In \citet{ayala:08} it was tested how sensitive the drop collisions are to the difference in Stokes' drag compared to real drag. Additional simulations with a non-linear drag formulation, which fits the measurements also for larger drops were performed. A decrease of up to 26~\% in the collision kernel for large drops was found and it was stated that no significant differences were found for small drops of up to 30~$\muup$m. However, the main results in \citet{ayala:08} were reported from simulations using Stokes' drag for drops in a size range of 10~to 60~$\muup$m. In order to produce comparable results, we use the same configuration in the simulations with ABC flow in this work.

Using Eq.~\ref{eq:fd} and adding gravity, the acceleration of a drop can be written as:
\begin{equation}
 \vec{a}=\frac{\partial \vec{v}}{\partial t}=\frac{9}{2}\frac{\rho_{\mathrm{a}}}{\rho_{\mathrm{w}}}\frac{\nu}{r^2}\Delta \vec{v}-\vec{g},
\label{eq:acc}
\end{equation}
with the time $t$, the drop density $\rho_{\mathrm{w}}$ and the gravitational acceleration vector $\vec{g}=(0,0,g')$, with $g'$ the reduced gravity of water drops in air which is essentially equal to the normal gravity.

For the numerical integration of Eq.~\ref{eq:acc}, a second order Verlet integrator was applied, which calculates the position and velocity at the next time step by:
\begin{eqnarray}
\label{eq:xverl}
 \vec{x}_{n+1}=\vec{x}_n+\vec{v}_n\Delta t+\frac{1}{2}\vec{a}_n\Delta t^2,\\
 \vec{v}_{n+1}=\vec{v}_n+\frac{1}{2}(\vec{a}_n+\vec{a}_{n+1})\Delta t,
\label{eq:vverl}
\end{eqnarray}
where $\vec{x}_n$ is the position, $\vec{v}_n$ the velocity and $\vec{a}_n$ the acceleration vector at time step $n$.

The position at the next time step can be calculated using Eq.~\ref{eq:acc} in Eq. \ref{eq:xverl}. In order to calculate the velocity at the next time step, Eq.~\ref{eq:acc} is substituted into Eq.~\ref{eq:vverl}. After solving for $\vec{v}_{n+1}$ this leads to the following equation:
\begin{equation}
 \vec{v}_{n+1}=\alpha_1\vec{v}_n+\alpha_2\left(\tau_{\mathrm{p}} \vec{a}_n+\vec{V}_{n+1}-\tau_{\mathrm{p}} \vec{g} \right),
 \label{eq:vnp1}
\end{equation}
with the flow velocity at the next time step $\vec{V}_{n+1}$. The factors $\alpha_1$ and $\alpha_2$ and the inertial time scale of the drop $\tau_{\mathrm{p}}$ are given by:
\begin{eqnarray}
 \alpha_1=\frac{1}{1+\frac{1}{2}\frac{\Delta t}{\tau_{\mathrm{p}}}},\\
 \alpha_2=\frac{\frac{1}{2}\frac{\Delta t}{\tau_p}}{1+\frac{1}{2}\frac{\Delta t}{\tau_{\mathrm{p}}}},\\
 \tau_{\mathrm{p}}=\frac{2}{9}\frac{\rho_{\mathrm{w}}}{\rho_{\mathrm{a}}}\frac{r^2}{\nu}.
\end{eqnarray}
Since the ABC flow is steady and the position at the next time step is known from Eq.~\ref{eq:xverl}, $\vec{V}_{n+1}$ is also known and Eq.~\ref{eq:vnp1} can be solved.

\subsection{Flow equations}
As an approximation for the turbulent flow in a cloud, the Arnold--Beltrami--Childress (ABC) flow \citep[e.g.][]{dombre:86} was implemented into the cloud model. ABC flow is an actual solution of the Euler equations unlike most forms of synthetic turbulence. It is defined by the following equations:
\begin{eqnarray}
 u=\alpha \left(A\sin(\tilde{z})+C\cos(\tilde{y})\right),
\label{eq:abc1}\\
 v=\alpha \left(B\sin(\tilde{x})+A\cos(\tilde{z})\right),\\
 w=\alpha \left(C\sin(\tilde{y})+B\cos(\tilde{x})\right),
\label{eq:abc3}
\end{eqnarray}
where $u$,$v$ and $w$ are the components of the velocity vector and $\tilde{x}$, $\tilde{y}$ and $\tilde{z}$ the components of the vector $\vec{\tilde{x}}=2\pi\cdot p \cdot \vec{x}/L$, with $\vec{x}$ being the position vector and $p$ the number of periods in the domain. The number of periods in the domain controls the length scale of the ABC flow $\lambda_{\mathrm{ABC}}$. For the coefficients $A$, $B$ and $C$ the following values were chosen: $A=1$, $B=\sqrt{2/3}$, $C=\sqrt{1/3}$. The factor $\alpha$ allows to scale the velocity of the ABC flow. For the given values of $A$, $B$ and $C$ it can be found that:
\begin{equation}
 u_{\mathrm{rms}}=\sqrt{2}\alpha,
\end{equation}
where $u_{\mathrm{rms}}$ is the root mean square velocity of the ABC flow.

\begin{figure}[th]
 \centerline{\includegraphics[width=0.5\textwidth]{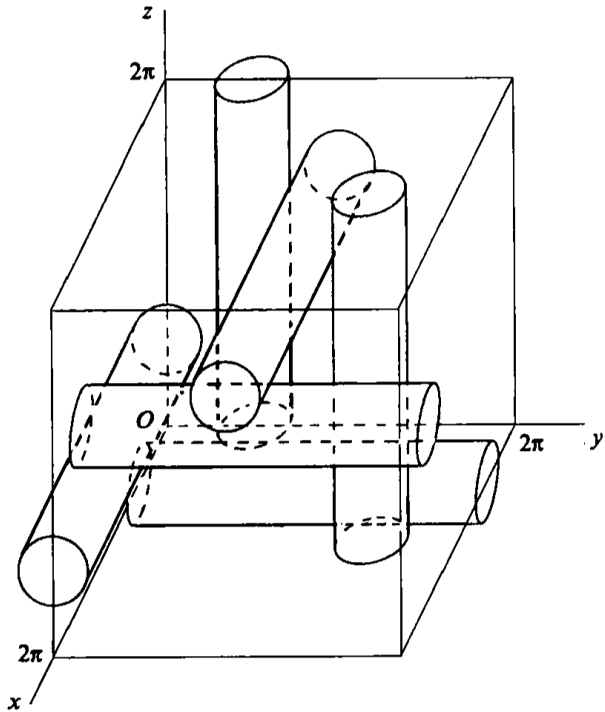}}
 \caption{Sketch of the vortices comprising the ABC flow \citep[taken from][]{dombre:86}.}
 \label{fig:abc_flow}
\end{figure}

Eqs.~\ref{eq:abc1} to \ref{eq:abc3} describe a number of vortex tubes (depending on $p$) in each spatial direction which are sketched in Fig.~\ref{fig:abc_flow} for the case of $p=1$. The flow field leads to chaotic advection for passive tracer particles. Note that drops are ballistic rather than passive, hence the requirement to have a prognostic equation for the acceleration of the drop.

The ABC flow does not reproduce the full complexity of unsteady atmospheric turbulence which consists of many different vortex scales whose kinetic energy typically follows the $k^{-5/3}$ law. But since cloud drops are mainly influenced by the smallest scales of the turbulent flow, the ABC flow might provide a good approximation for the simulation of the movement of cloud drops, if we choose the scales of the ABC flow to be commensurate with the smaller scales in observed turbulence.

The characteristics of the ABC flow can be controlled by setting the root mean square velocity $u_{\mathrm{rms}}$ and the length scale $\lambda_{\mathrm{ABC}}$ through $p$ and $\alpha$. It is assumed that $\lambda_{\mathrm{ABC}}$ should be close to the Taylor microscale of the flow which is to be simulated. Similarly, $u_{\mathrm{rms}}$ of the ABC flow should be close to the one of the studied case. The choice of the Taylor microscale and root mean square velocity seems practical, since these two parameters are linked to the dissipation rate and might allow to adjust the characteristics of the flow to different real turbulence cases. Since the characteristics of the ABC flow are different from real turbulence, these parameters might have a different effect on the drop collisions compared to real turbulence. To get more insight into this problem a sensitivity study is carried out in section \ref{sec:results} to investigate the influence of the ABC flow parameters on the collision statistics of the drops.

\subsection{Collision detection}

Detecting the collisions of cloud drops in the model domain is a numerically expensive task. The numerical effort of finding collisions scales with $N^2$, where $N$ is the number of drops, if every pair of drops is considered. In order to save computational time, a hash table approach called the linked-cell method \citep[e.g.][]{allen:17,welling:11} was used. In this method the model domain is divided into subdomains and at each time step a list is produced which records which drop is located in each subdomain. For detecting collisions, only drops in the same subdomain or neighbouring subdomains are considered. For the partitioning of the model domain it is important to consider the maximum distance a drop can travel within a time step. This method scales with $N$ and is therefore much more efficient especially for large numbers of drops (in our simulations we typically simulated trajectories for 250000~drops).

The collision detection mechanism used in this work registers a collision in two cases. The first case occurs simply when the distance between two drops in the current time step is less than the sum of their radii. For the second case it is taken into account that collisions can also happen in between time steps. To detect such collisions a simple linear approach for the motion of two drops, which were sufficiently close to each other in the previous time step, is applied in between time steps. The linear equation of drop motion can be substituted into the equation for the distance between the drops and the resulting quadratic equation can be solved for time. If the time at which two drops reach a distance closer than the sum of their radii is within the current time interval a collision is detected.

On collision one of the colliding drops is re-positioned to a random location in the model domain under the condition that the drop is at a minimum distance to any other drop at the new location. This technique allows the number of drops to remain constant after collision so as to build up reliable collision statistics for a particular drop population. The effect of the random displacement of one of the colliding drops has been tested across various test cases and was not found to influence collision statistics substantially.

The motion of heavier drops tends to be dominated by the terminal fall speed, moving mostly vertically with a small perturbation due to turbulence. If the domain was treated fully triply periodic, this would mean that heavier drops get essentially frozen into their horizontal location, and thus stop contributing reliably to collision statistics. To break this effect of frozen-in horizontal locations for heavy drops, drops get shifted randomly in the horizontal plane when they reach the lower periodic boundary of the domain. The distance of displacement is proportional to the size of the drop, in order to focus this method on the heaviest drops. The horizontal displacement of falling drops mimics the effect of drops from other parts of the cloud falling into the domain and drops leaving the domain respectively, while keeping the drop size spectrum constant.

We have tested the effect of this horizontal displacement by comparing our model with theoretical kernels for drops that fall uniformly, without turbulence, with their terminal velocities. Here, the displacement algorithm is in fact essential to produce reliable collision statistics because in this case the horizontal locations are rigorously frozen in for all non-colliding drops. With the displacement algorithm, we can reproduce the theoretical kernels, except for collisions between similar sized small drops (around 10~$\muup$m) where the expected collision rate is very small and numerical estimates suffer from a lack of sampling.

\subsection{Collision kernel}

The collision kernel describes the ability of drops of different sizes to collide with each other. To calculate the collision kernel from the numerical simulations presented in this work, the number of collisions between two different size classes of drops $N_{12}$ is normalised by the number of drops in each size class $N_i$ (with $i$ being the index of the size class), the simulation time $t$ and the domain volume $V$:
\begin{equation}
 \Gamma_{12}^{\mathrm{D}}=\frac{N_{12}V}{N_1N_2t}.
\end{equation}
$\Gamma_{12}^{\mathrm{D}}$ is called the dynamic collision kernel.

An alternative description of drop collisions is given by the kinematic collision kernel. If aerodynamic interactions are not considered, it is equal to the dynamic collision kernel. In the case of drops falling with their respective terminal velocities in still air it can be written as:
\begin{equation}
 \Gamma_{12}^{\mathrm{g}}=\pi R^2 |v_1-v_2|,
\end{equation}
with $R = r_1+r_2$ being the sum of the radii of the two colliding drops and their terminal velocities $v_1$ and $v_2$.

Considering the aerodynamic interaction of drops and the influence of the turbulent flow, the equation for the collision kernel can be written as \citep{wang:05}:
\begin{equation}
 \Gamma_{12}=\eta_{\mathrm{E}}\,\eta_{\mathrm{G}} \,E_{12} \,\Gamma_{12}^{\mathrm{g}},
\end{equation}
with $E_{12}$ being the collision efficiency which considers the aerodynamic interaction of drops, the enhancement factor for the collision efficiency due to turbulence $\eta_{\mathrm{E}}$ and the enhancement factor for the collision kernel due to turbulence $\eta_{\mathrm{G}}$.

Since aerodynamic interactions between drops are not considered in the present model, $E_{12}$ and $\eta_{\mathrm{E}}$ are equal to~1 and $\Gamma_{12}=\Gamma_{12}^{\mathrm{D}}$. Therefore, the enhancement factor due to the turbulent flow can be calculated by:
\begin{equation}
 \eta_{\mathrm{G}}=\frac{\Gamma_{12}^{\mathrm{D}}}{\Gamma_{12}^{\mathrm{g}}}.
 \label{eq:eps}
\end{equation}
A normalised version of the collision kernel is used in this work for which the dynamic collision kernel is divided by the contribution of the drop sizes to the kernel:
\begin{equation}
 \Gamma_{12}'=\frac{\Gamma_{12}^{\mathrm{D}}}{\pi R^2}.
\end{equation}
This normalised kernel has units of velocity and can be interpreted as an average effective velocity difference between the colliding drops. It removes the quadratic increase of the kernel with the sum of the drop radii and therefore makes it easier to see the details in different collision kernels. For simulations without turbulence we verified that $\Gamma_{12}'$ equals the difference in terminal velocities.

\subsection{Model setup}

The model domain is 10$\times$10$\times$10~cm cubic with periodic boundaries in all three directions. Initially, the domain was populated by 250000 randomly distributed spherical drops. Due to the efficiency of the ABC flow simulations, such a large number of drops can be simulated easily on standard hardware. The simulations were conducted on a single core of a 24-core Intel Xeon processor with 2.3~GHz. It took around 35~h to perform a simulation of 30~min (750 eddy turnover times) with a time step size of $10^{-3}$~s. Sensitivity tests with different time steps have been conducted to ensure that the chosen time step resolves the drop motion and collisions sufficiently.

For each simulation a spin up run of 25 eddy turnover times was performed, in order to allow the drops to redistribute dynamically in a way that is consistent with the prescribed flow. After that, another 750 eddy turnover times were simulated to gather the collision statistics. Two kinds of simulations were conducted. The first kind of simulation was set up with a low size resolution with only six drop size bins: 10, 20, 30, 40, 50 and 60~$\muup$m. The drops were equally distributed in the six size bins.

For the second kind of simulation a high size resolution with 50 size bins in the interval between 10 and 60~$\muup$m was used. To compensate for the fact that large drops collide much more often than small drops, a log-normal size distribution was applied for the runs with high size resolution. That means that there were more small drops than large drops in the domain which increases the number of collisions and improves the collision statistics for small drops.

To assess the quality of the model, several runs with different configurations were run, and the results were compared to the DNS results by \citet{ayala:08}. Table \ref{tab:setup} gives an overview over the parameters used in the \citet{ayala:08} simulation.

\begin{table*}[]
 \begin{center}
 \begin{tabular} {c|c|c|c|c}
  $\rho_{\mathrm{w}}$ & $\rho_{\mathrm{a}}$ & $\nu$ & $\Delta$ t & L \\
  \hline
  1000~kg m$^{-3}$ &  1~kg m$^{-3}$ &  0.17 cm$^2$ s$^{-1}$ & 0.001~s &  10 cm$^3$ \\
 \end{tabular}
 \end{center}
 
 \begin{center}
 \begin{tabular} {c|c|c|c|c}
  $\epsilon$ & $\lambda$ & u$_{\mathrm{rms}}$ & L$_{\mathrm{f}}$ & $\eta$\\ 
  \hline
  400~cm$^2$ s$^{-3}$ &  1.0~cm &  12.42~cm s$^{-1}$ &  6.8~cm & 8.4$\times 10^{-2}$~cm
 \end{tabular}  
 \end{center}
 \caption{Model parameters in the DNS by \citet{ayala:08}: density of water $\rho_{\mathrm{w}}$, density of air $\rho_{\mathrm{a}}$, kinematic viscosity $\nu$, time step size $\Delta t$, edge length of the cubic domain $L$, dissipation rate $\epsilon$, Taylor microscale $\lambda$, root mean square velocity $u_{\mathrm{rms}}$, integral length scale $L_{\mathrm{f}}$ and Kolmogorov scale $\eta$.}
 \label{tab:setup}
\end{table*}

The sensitivity of collision statistics to changes in the ABC flow was studied by conducting 7 simulations with low drop size resolution and different length scales and flow velocities (see Tab.~\ref{tab:ABC}). In the names of the simulation cases the number after ABC refers to the number of periods for the sine and cosine in the ABC flow equations (Eqs.~\ref{eq:abc1}--\ref{eq:abc3}). For ABC1 one period fits in the model domain in each direction. Since one period consists of two vortex tubes, the diameter of one vortex is half the edge length of the cubic domain $L$. Therefore the ABC length scale is 5\,cm in the ABC1 case. In the last three lines of Tab.~\ref{tab:ABC} the number after the comma refers to the scaling of the RMS velocity. E.g. ABC5,0.2u is a case where 5 ABC periods fit in the model domain in each direction and the RMS velocity of the flow is 0.2~times the RMS velocity of the ABC5 case.

\begin{table}[t]
 \begin{center}
 \begin{tabular} {l|r|r}
  Case & $\lambda_{\mathrm{ABC}}$ & $u_{\mathrm{rms}}$\\
  \hline 
  ABC1 & 5~cm &  12.42~cm s$^{-1}$ \\
  ABC2 & 2.5~cm & 12.42~cm s$^{-1}$\\
  ABC5 & 1~cm & 12.42~cm s$^{-1}$\\
  ABC10 & 0.5~cm & 12.42~cm s$^{-1}$\\
  ABC5,0.2u & 1~cm & 2.48~cm s$^{-1}$\\
  ABC5,0.5u & 1~cm & 6.21~cm s$^{-1}$\\
  ABC5,2u & 1~cm & 24.84~cm s$^{-1}$
 \end{tabular}
 \end{center} 
 \caption{ABC flow parameters for sensitivity test cases with low size resolution.}
 \label{tab:ABC}
\end{table}

\section{Results}
\label{sec:results}

Results of simulations with low size resolution, using the \citet{ayala:08} case settings are presented for different ABC length scales (Sec.~\ref{sec:sens_l}) and flow velocities (Sec.~\ref{sec:sens_u}). The results are compared to the results by \citet{ayala:08} for the eddy dissipation rate $\epsilon=400$~cm$^2$~s$^{-3}$ and Reynolds number $R_{\lambda}=72.4$. 

\subsection{Sensitivity to different length scales}
\label{sec:sens_l}

\begin{figure*}[]
 \centerline{\rotatebox{-90}{\includegraphics[height=0.70\textwidth]{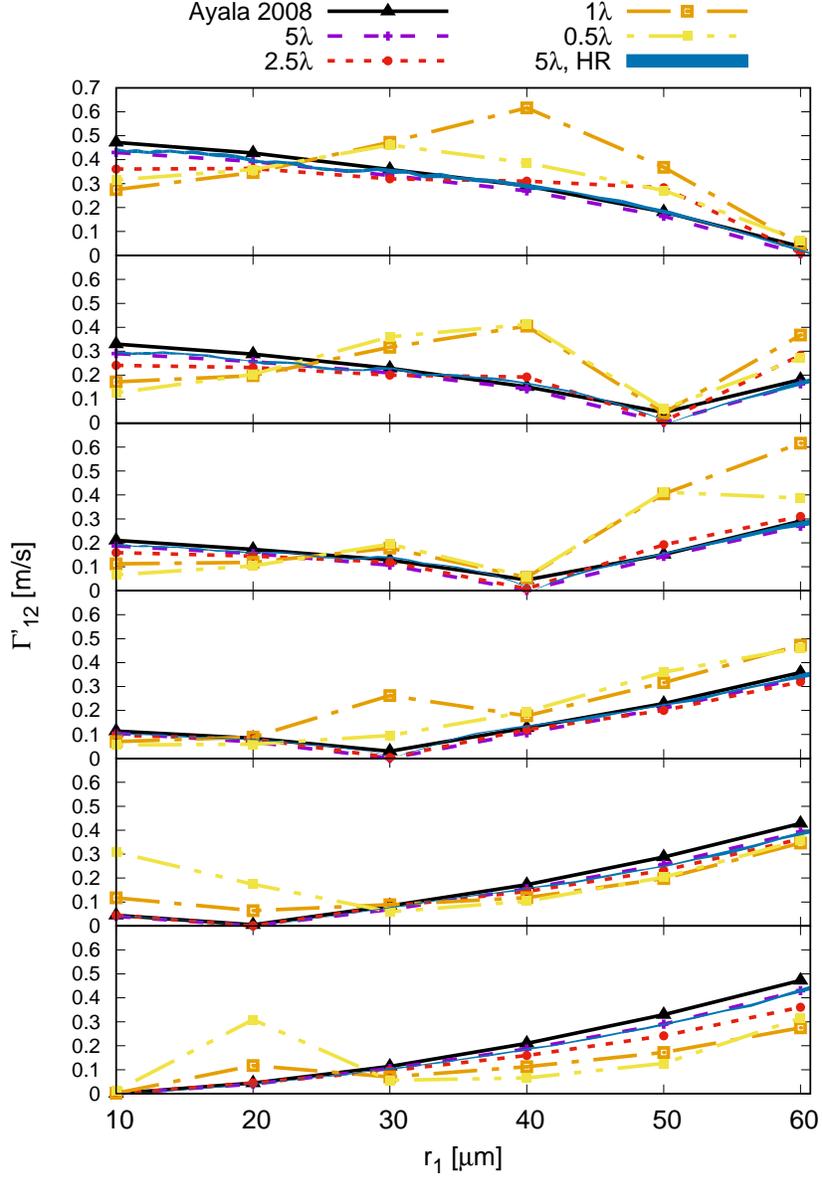}}}
 \caption{Normalised collision kernels $\Gamma'_{12}$ in m~s$^{-1}$ from the DNS of \citet{ayala:08} (black, solid), the simulation with an ABC length scale $\lambda_{\mathrm{ABC}}$ of $0.5$ times the Taylor microscale $\lambda$ (yellow, dash-dot-dot), $\lambda_{\rm ABC}=\lambda$ (orange, dash-dot), $\lambda_{\rm ABC}=2.5\lambda$ (red, dot), $\lambda_{\rm ABC}=5\lambda$ (purple, short dash) and the envelope of 11 realisations of the high-resolution simulation for $\lambda_{\rm ABC}=5\lambda$. The kernels are given for drop radii $r_2$ of 10, 20, 30, 40, 50 and 60~$\muup$m (from bottom to top panel).}
 \label{fig:kernels_ay}
\end{figure*}

In the first set of simulations different length scales were used while all other parameters were kept constant, including the RMS velocity which was set to be equal to the one in \citet{ayala:08}. The results of these simulations are shown in Fig.~\ref{fig:kernels_ay}. It shows the normalised kernels for collisions of 10, 20, 30, 40, 50 and 60~$\muup$m drops, respectively (from bottom to top). The kernels from the DNS reported by \citet{ayala:08} (black solid line) show the typical structure for falling drops, where a minimum occurs for same sized drops \citep[e.g.][]{siewert:14}. This is due to the relative velocities between same sized drops being very small.

It can be seen that the kernels of the cases with the two shortest length scales ($\lambda_{\mathrm{ABC}}=0.01$~m and $\lambda_{\mathrm{ABC}}=0.005$~m) are not capturing the structure of the kernels from the DNS very well. For 10, 20 and 30~$\muup$m they fail to show the minimum for same sized drops and in general over or under predict the normalised kernel. As the ABC length scale increases, the normalised kernels fit better with the DNS results. For the largest length scale ($\lambda_{\mathrm{ABC}}=0.05$~m) the deviations from the DNS kernels are the least over the whole size range. However, in that case the normalised kernel for same sized drops is significantly smaller than in the DNS. The kernels for the high resolution simulations will be discussed in Section~\ref{sec:full}.

In order to get a quantitative measure of how much the ABC flow simulations and the DNS agree the following metric is used:
\begin{equation}
 q = \exp\left( \overline{ \log^2\left({\Gamma_{12}}/{\Gamma_{12}^{\mathrm{a}}}\right) } \right)^{1/2},
 \label{eq:q}
\end{equation}
where $\Gamma_{12}^{\mathrm{a}}$ is the collision kernel from the DNS. The overbar denotes the average over all size bins. Essentially, $q$ is the standard deviation of all the ratios between the collision kernel values, where the logarithm allows for equitable contributions between ratios smaller than one and larger than one.

\begin{figure}[]
 \begin{center}
  \rotatebox{-90}{\includegraphics[height=0.45\textwidth]{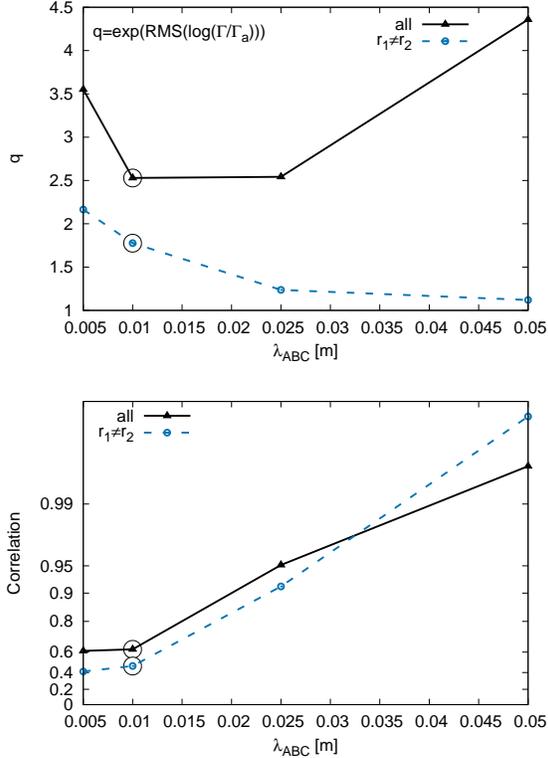}}
 \end{center}
 \caption{Top Panel: $q$-Metric for the deviation of the ABC flow simulations from the DNS results for the collision kernels for different ABC length scales (black, solid), and omitting contributions from same-sized drops (blue, dashed). Circles mark the values for the case where the ABC length scales is equal to the Taylor microscale of the DNS. Bottom panel: as above, but for the correlation between the kernels across the drop size spectrum. The $y$-axis is scaled by Fisher's transformation for correlation coefficients.}
 \label{fig:l_eqi}
\end{figure}

Results for $q$ and the correlation between the kernels from the DNS and ABC simulations for different ABC length scales are presented in Fig.~\ref{fig:l_eqi}. Since the kernels for same sized drops are very small but sometimes deviate significantly from the DNS kernel, $q$ and the correlations were also calculated by excluding the kernels for same sized drops. Results for $q$ show that the collision kernels from the ABC flow simulations are in best agreement with the DNS if the ABC length scale is 2.5 times the Taylor microscale from the DNS ($\lambda_{\mathrm{ABC}}=0.025$~m). For an ABC length scale of 5 times the Taylor microscale ($\lambda_{\mathrm{ABC}}=0.05$~m) the match with the DNS is worse, due to the contribution of same-sized drop collisions; if same sized drop collision kernels are omitted, the results for this length scale fit the best. 

The results for the correlations in Fig.~\ref{fig:l_eqi} show that for larger ABC length scales the correlation is increasing with its largest value for the largest length scale. This is true for correlations including and excluding same size drop collision kernels.

These results indicate that the structure of the DNS collision kernel is best captured by the simulation with an ABC length scale of 5 times the Taylor microscale ($\lambda_{\mathrm{ABC}}=0.05$~m).

\subsection{Sensitivity to different flow velocities}
\label{sec:sens_u}

\begin{figure*}[]
 \centerline{\rotatebox{-90}{\includegraphics[height=0.70\textwidth]{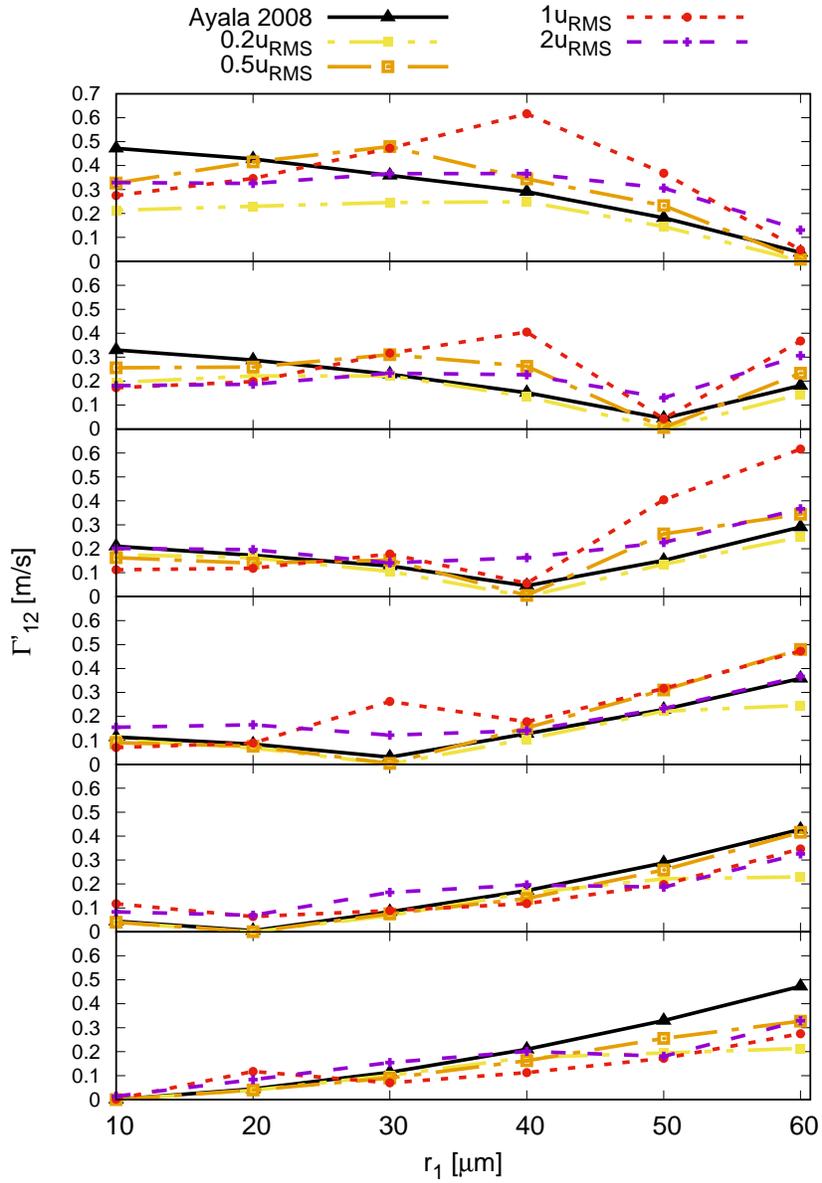}}}
 \caption{Normalised collision kernels $\Gamma'_{12}$ in m~s$^{-1}$ from the DNS of \citet{ayala:08} (black, solid), the simulation with an RMS velocity in the ABC flow $u_{\mathrm{rms}}$ of 0.2 times the RMS velocity of the DNS $u'$ (yellow, dash-dot-dot), $u_{\mathrm{rms}}=0.5u'$ (orange, dash-dot), $u_{\mathrm{rms}}=u'$ (red,dot) and $u_{\mathrm{rms}}=2u'$ (purple, short dash). The kernels are given for drop radii $r_2$ of 10, 20, 30, 40, 50 and 60~$\muup$m (from bottom to top panel).}
 \label{fig:kernels_ay_u}
\end{figure*}

To test the sensitivity of the results to a change of flow velocity, four simulations with different RMS velocities  of the ABC flow were conducted. The ABC length scale in all simulations was kept constant at $\lambda_{\mathrm{ABC}}=1$~cm, the length scale of the ABC5 case. The RMS velocity of the ABC flow was set to $2.48$~cm~s$^{-1}$, $6.21$~cm~s$^{-1}$, $12.42$~cm~s$^{-1}$ (the RMS velocity of the DNS), and $24.84$~cm~s$^{-1}$. The collision kernels resulting from these simulations can be seen in Fig.~\ref{fig:kernels_ay_u}.

In the case of the largest flow velocity (2 times the RMS velocity of the DNS) the dependence of the collision kernel on the drop size is generally smaller than in the DNS. Decreasing the ABC velocity scale by half leads to an increased dependency of the collision kernel on the drop size but large deviations from the DNS can still be seen for most drop sizes. A clear improvement in matching the DNS results is visible when the velocity in the ABC flow is further decreased.

\begin{figure}[]
 \begin{center}
  \rotatebox{-90}{\includegraphics[height=0.45\textwidth]{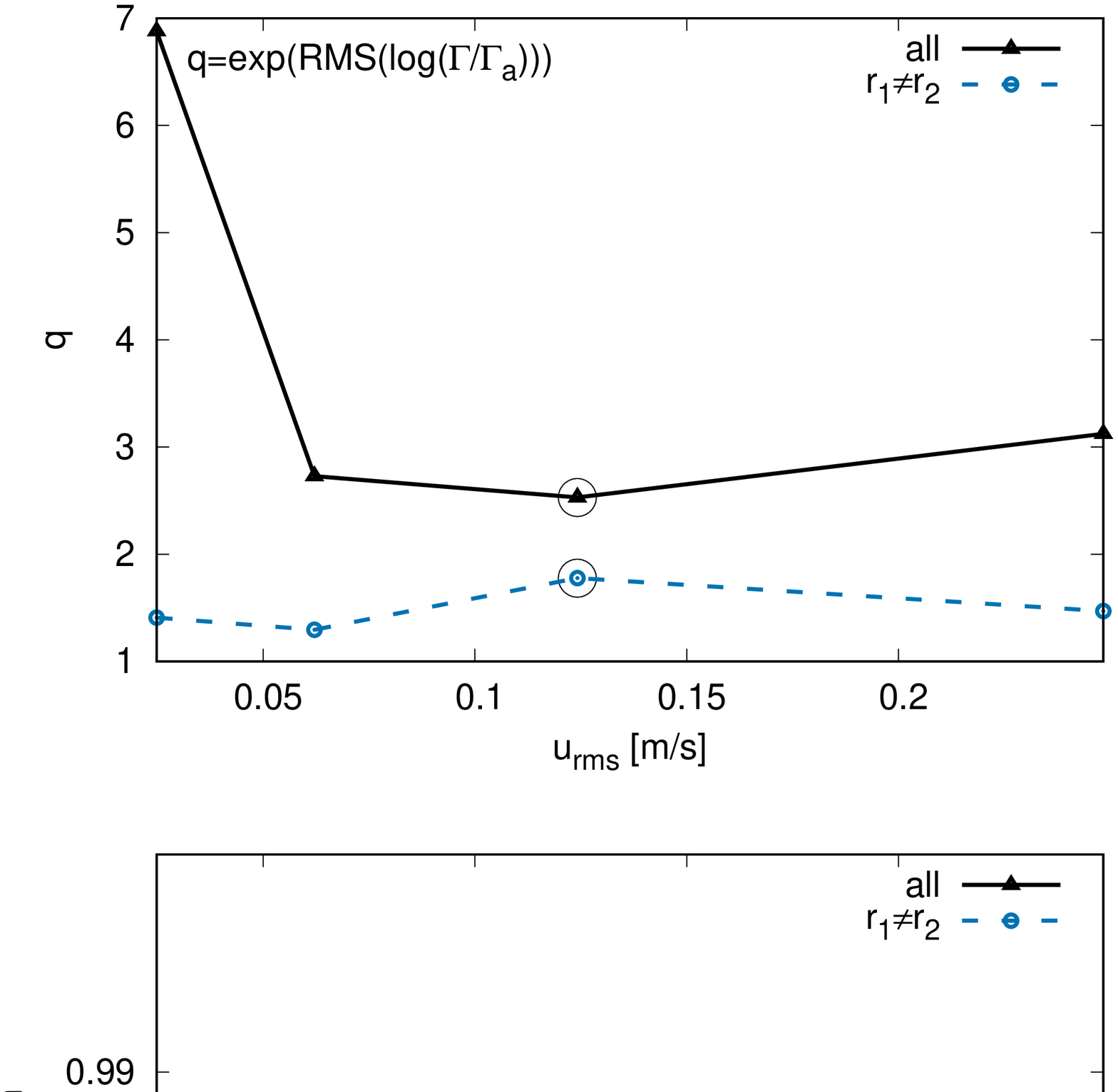}}
 \end{center}
 \caption{Top panel: $q$-Metric for the deviation of the ABC flow simulations from the DNS results for the collision kernels for different RMS velocities of the ABC flow (black, solid), and omitting contributions from same-sized drops (blue, dashed).  The circles mark the values for the case where the RMS velocity of the ABC flow is equal to the RMS velocity of the DNS. Bottom panel: as above, but for the correlation between the kernels across the drop size spectrum. The $y$-axis is scaled by Fisher's transformation for correlation coefficients.}
 \label{fig:urms_eqi}
\end{figure}

To quantify the effect of a change of RMS velocity on the collision kernel the metric from Eq.~\ref{eq:q} is used again. The results are shown in Fig.~\ref{fig:urms_eqi}. For smaller RMS velocities the deviation between DNS and ABC flow is less. Except for the smallest RMS velocity where a strong increase in deviation can be seen. From the curve in which collisions between equal sized drops are ignored, it is obvious that this increase is caused by deviations for equal sized drops. For collisions of drops of unequal size the dependency of the deviation from the DNS with RMS velocity is low. However, for smaller RMS velocities the structure of the DNS collision kernel is reproduced better, as can be seen in the correlation coefficient in the bottom panel of Fig.~\ref{fig:urms_eqi}.

The better fit with the DNS results for smaller RMS velocities indicates that a smaller centrifugal force (since the length scale was kept constant) acting on the drops leads to a better agreement. The same behaviour could be seen in the results of the simulations with different length scales. A larger length scale, meaning a smaller centrifugal force (due to the larger radius of the ABC vortices while keeping the RMS velocity constant), leads to a better agreement between the DNS and ABC flow results. Given the same length scale and RMS velocity the centrifugal force produced by the ABC flow vortices is too large compared to the average effect of a complex DNS turbulence flow. But by reducing the centrifugal force, especially by increasing the ABC length scale, similar results to the DNS can be achieved.

\subsection{Kernel with high drop size resolution}
\label{sec:full}

In this section results from simulations with high drop size resolution are presented. These runs will be referred to as HR runs. For the HR simulations the RMS velocity of the ABC flow was set to 12.42~cm~s$^{-1}$ (the RMS velocity of the DNS by \citet{ayala:08}) and the ABC length scale was set to 5 times the Taylor microscale in the DNS (the settings of the ABC1 run, see Tab.~\ref{tab:ABC}), since these yield the best results. The simulations use 50 size bins, as opposed to 6 bins in the low size resolution simulations (hereafter referred to as LR runs) and the DNS of \citet{ayala:08}. The results of the HR runs are compared to the LR run with the same parameter setting of the ABC flow (ABC1).

To improve the statistics of the HR results, averages over 11 runs were calculated. However, since the variance in the kernels for those 11 realisations is small, the effect compared to the single results is limited, indicating that the single results are already sufficiently representative.

Like in the LR runs the total number of drops in the HR runs is 250000, leading to an average of 5000 drops per bin. Since smaller drops are experiencing fewer collisions than larger drops, a log-normal size distribution was chosen with larger numbers of smaller drops and smaller numbers of larger drops. The smallest sizes (10~$\muup$m) have about 9500 drops per bin while the largest sizes (60~$\muup$m) have about 500 drops per bin.

As a consequence of the logarithmic distribution of size bins, the drop sizes of the LR runs do not exactly coincide with the drop sizes of the HR runs. When comparing the results, the next closest size bin to the size bin of the LR runs is chosen (see Tab.~\ref{tab:size}). Since the deviation of the HR size bins from the LR size bins is very small it is assumed that this does not affect the comparison of the results. For simplicity, when comparing the two simulations, the size bins of the HR runs will be referred to by their LR equivalent.

\begin{table*}[t]
 \begin{center}
 \begin{tabular} {c|c|c|c|c|c|c}
  LR & 10 & 20 & 30 & 40 & 50 & 60\\ 
  \hline
  HR & 10 & 20.126 & 30.172 & 40.504 & 50.515 & 60.723
 \end{tabular}  
 \end{center}
 \caption{Drop sizes (in $\muup$m) used for comparison of results from the low resolution (LR) and high resolution (HR) simulations.}
 \label{tab:size}
\end{table*}

The blue line in Fig.~\ref{fig:kernels_ay} shows the envelope of the collision kernels of all 11 HR runs. It is very similar to the LR run with the same length scale ($\lambda_{\mathrm{ABC}}=5\lambda$) and also very close to the DNS results.

\begin{figure}[t]
 \centerline{\rotatebox{-90}{\includegraphics[height=0.5\textwidth]{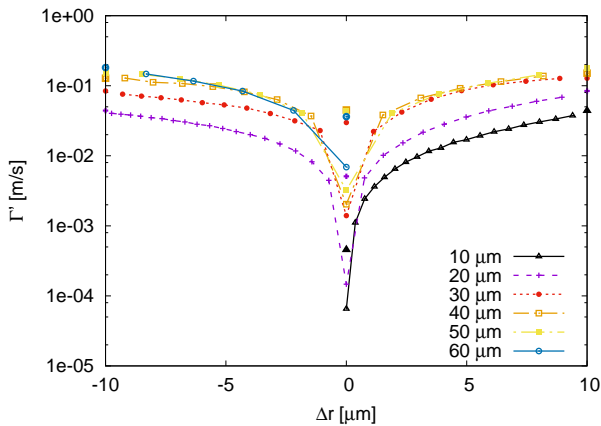}}}
 \caption{Normalised collision kernels $\Gamma'_{12}$ in m~s$^{-1}$ from the DNS (symbols) and averaged over the runs with high drop size resolution (lines with symbols) for different differences in radius $\Delta r$ of collising drops in $\muup$m.}
 \label{fig:same}
\end{figure}

Due to the higher resolution of the HR run it is possible to have a closer look at the deviations of the kernel for same sized drops. Fig. \ref{fig:same} shows the averaged kernels from the HR runs (lines with symbols) for 6 different drop radii $r_1$ for collisions with drops in a radius range of $-10< \Delta r<10$ $\muup$m (with $\Delta r$ being the difference between the radii of the colliding drops) and compares it to the values of the DNS (symbols). Since the DNS data is not available with high resolution an exact comparison is not possible and values for the collision kernels of the DNS can only be shown for a difference in radius of -10, 0 and 10 $\muup$m. But it can be seen that the large deviations for similar sized drops occur most likely in a narrow range of about $2\Delta r$ around 0 $\muup$m where a sharp drop of collision kernels aoccur. Outside of this interval the kernels do not exhibit such a strong decrease of collision kernels and the values are closer to the trend of the DNS values.

\begin{figure}[t]
 \centerline{\rotatebox{-90}{\includegraphics[height=0.6\textwidth]{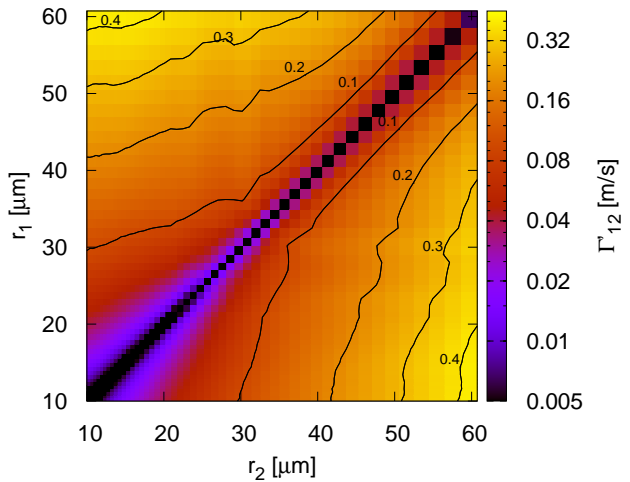}}}
 \caption{Normalised collision kernels $\Gamma'_{12}$ in m~s$^{-1}$ averaged over the runs with high drop size resolution (colours and contour lines). The kernel is given for collisions of drops with radius $r_1$ and $r_2$ in $\muup$m.}
 \label{fig:kernels_full}
\end{figure}

Figure~\ref{fig:kernels_full} shows the normalised collision kernel from the run with high drop size resolution. It shows that same sized drops experience significantly fewer collisions than collision partners with larger differences in size. It can also be seen that the gradient at smaller drop sizes is smaller, meaning that collisions of small drops among each other are less likely.

\begin{figure}[t]
 \centerline{\rotatebox{-90}{\includegraphics[height=0.6\textwidth]{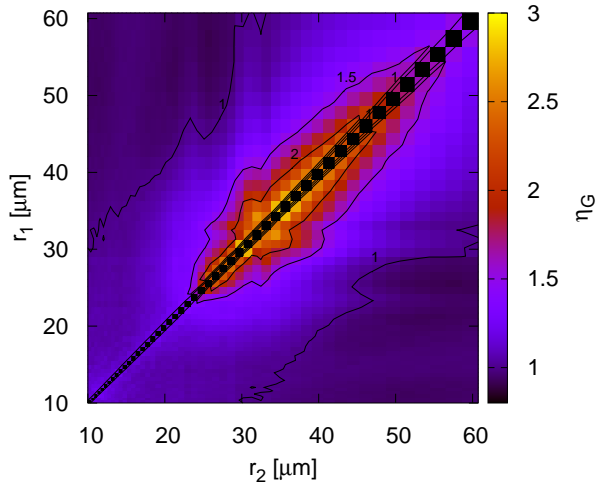}}}
 \caption{Enhancement factor $\eta_{\mathrm{G}}$ for the average over the runs with high drop size resolution (colours and contour lines). The enhancement factor is given for collisions of drops with radius $r_1$ and $r_2$ in $\muup$m.}
 \label{fig:eff2d}
\end{figure}

The normalised collision kernel represents the combined effect of the turbulent flow and the difference in terminal velocities due to the different drop sizes. In order to isolate the effect of the turbulent flow the enhancement factor for the collision kernel $\eta_G$ (see Eq.~\ref{eq:eps}) is plotted in Fig.~\ref{fig:eff2d}. It shows that the main effect of the turbulent flow is to enhance the collision kernel of similar sized drops. The influence on collisions of other drops is very small and the enhancement factor is in general decreasing with smaller drop sizes. This is also in line with DNS results presented by \citet{siewert:14}.

\begin{figure*}[t]
 \centerline{\rotatebox{-90}{\includegraphics[height=1\textwidth]{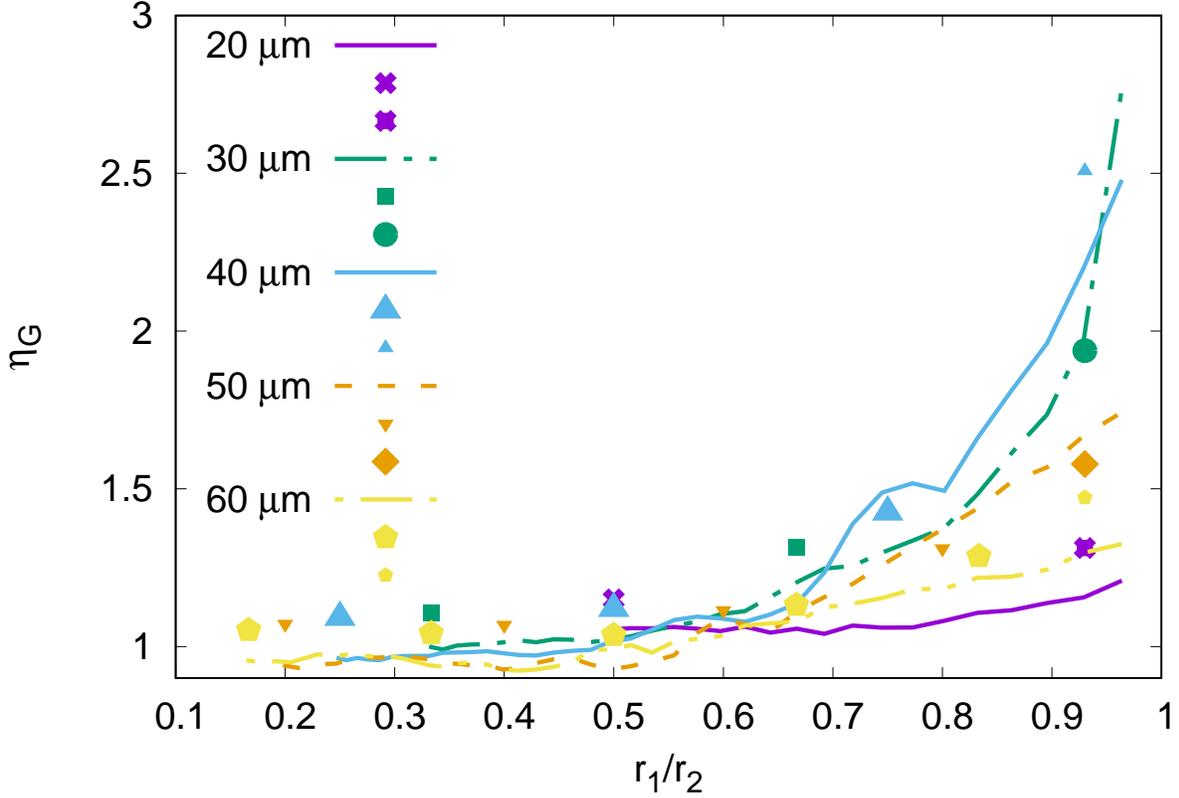}}}
 \caption{Enhancement factor $\eta_{\mathrm{G}}$ over drop size ratio $r_1/r_2$ ($r_2$ is the larger drop) from the simulation with high drops size resolution (lines) for the different sizes of the larger drop given in the legend. The symbols are the corresponding enhancement factors from the DNS from \citet{ayala:08}. The symbols at $r_1/r_2=0.93$ are from bimodal simulations.}
 \label{fig:eff}
\end{figure*}

A similar picture is drawn in Fig.~\ref{fig:eff}, where the enhancement factor is plotted over the drop size ratios of colliding drops. The lines in the figure represent the average over the results from the HR runs. The symbols are the results from the DNS by \citet{ayala:08}, except for the symbols at $r_1/r_2=0.93$ which were produced by additional runs with only 2 drop sizes per run in order to add data points at a larger drop size ratio.

For small drop size ratios the results of the HR runs and the DNS both show little influence of the turbulence on drop collisions. There is only a small increase in the enhancement factor up until drop size ratios of 0.6. In this range the enhancement factor is very similar for the same drop size ratio, independently of the actual drop size. However the DNS results in this size ratio range are larger than the HR results. The average enhancement factor in the interval up to $r_1/r_2=0.6$ is 0.98 in the HR simulations while in the DNS the average $\eta_{\mathrm{G}}$ in the same interval is 1.08, indicating a slight enhancement of collisions due to the turbulence in that range. This effect is not reproduced in the HR simulation where a slight decrease of collisions is seen due to the turbulence in that size ratio range.

For larger ratios $\eta_{\mathrm{G}}$ is similar in the DNS and HR simulation. Even though it is difficult to compare the enhancement factor for larger size ratios due to the lower drop size resolution of the DNS.

Since for larger size ratios (similar sized drops) the number of collisions in the HR run are relatively small, additional runs with only two sizes of drops (bimodal runs with a size ratio of 0.93) have been conducted to confirm that the shape of the enhancement factor for the HR simulation is not corrupted by a small sample size for large size ratios.

The enhancement factor for the 20~$\muup$m drop is almost constant over the whole interval and only shows a small increase for $\eta_{\mathrm{G}}$ larger than 0.8. The value from the bimodal run is slightly larger than in the HR runs. There is a moderate increase for the 60~$\muup$m drop in the HR run with values at drop size ratio 0.93 of $\eta_{\mathrm{G}}=1.5$. The enhancement factor in the bimodal run is larger with $\eta_{\mathrm{G}}=1.9$. Values for 30~$\muup$m and 50~$\muup$m are very close to each other just below $\eta_{\mathrm{G}}=2$ in both the HR and bimodal run. The largest increase can be seen for the 40~$\muup$m drop with values around 2.5 at $r_1/r_2=0.93$ in both the HR and bimodal run.

For larger drop size ratios a spread in $\eta_{\mathrm{G}}$ occurs, which is increasing with larger ratios. In the HR runs the spread starts to occur at around $r_1/r_2=0.7$. There are only 3 data points in the DNS results larger than 0.7 but, they seem to indicate that such spreading also occurs in the DNS data with the value for 40~$\muup$m drops increasing stronger than for the 50 and 60~$\muup$m drops. The results from the additional bimodal runs are very similar to the HR runs and confirm that the spread is not a result of insufficient collision numbers in the HR runs. The LR runs produced very similar enhancement factors to the HR runs and are not shown here.

\begin{figure}[t]
 \centerline{\rotatebox{-90}{\includegraphics[height=0.5\textwidth]{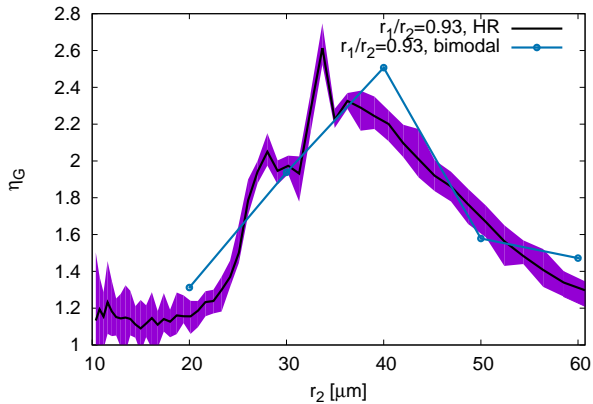}}}
 \caption{Enhancement factor $\eta_{\mathrm{G}}$ over drop size $r_2$ for the drop size ratio $r_1/r_2=0.93$ from the HR (black) and bimodal runs (blue with circles).}
 \label{fig:bieta}
\end{figure}

The enhancement factor at $r_1/r_2=0.93$ is plotted in Fig.~\ref{fig:bieta}. For the bimodal runs, the maximum occurs for collision pairs with $r_1=37.2$~$\muup$m and $r_2=40$~$\muup$m and for the HR runs for $r_1=31.3$~$\muup$m and $r_2=33.7$~$\muup$m. Due to the low size resolution of the bimodal runs it is not possible to compare the exact location of the peaks. The narrow envelope of the HR runs indicates that much of the structure in the turbulent enhancement factor does not result from sampling noise.

\section{Conclusions}
\label{sec:conc}

DNS is the most accurate available technique to simulate the effect of turbulent flow on drops; DNS is however a computationally very expensive method. In this work it was shown that, when only gravity and the drag on the drops is considered, similar results to DNS can be achieved by using a simple ABC flow instead of using fully resolved turbulence. 

Two different types of simulations were conducted: low drop size resolution (LR) simulations with the same 6 size bins that were used in the reference DNS from \citet{ayala:08} and high drop size resolution (HR) runs covering the same size interval but using 50 bins instead. In both cases 250000 drops were released into the flow.

In the LR runs it was shown that there is a good match in collision kernels between the DNS and the ABC simulation. As expected, we find that the collision kernels depends on the choice of turbulence parameters. The results show that the match with the DNS improves if the ABC vortices are larger (in this case 5 times larger) than the Taylor microscale in the DNS or the RMS velocity is chosen smaller (in this case 2 times smaller) than the RMS velocity in the DNS. While the length scale had to be chosen larger than the Taylor microscale in the DNS when keeping the RMS velocity constant, it was also shown that there was not much difference in the results when setting the ABC flow length scale to $2.5\lambda$ or $5\lambda$. A relatively big range of length scale settings yielded a good match with the DNS data.

We speculate that in general the effects of the centrifugal force acting on the drops due to the ABC flow is stronger than in the DNS if the Taylor micro scale is chosen as vortex diameter and the RMS velocity is equal in the ABC and DNS flow. The ABC flow is different from the DNS: it consists of only one size of stationary vortices. In a DNS study \citet{franklin:05} found that stationary flow fields produce larger collision kernels because the drops are exposed to the same structures for a longer time. As a consequence the radial relative velocities of the drops were larger in the stationary flow field than in the unsteady DNS. In future studies an unsteady version of the ABC flow could be attempted to assess the influence of this effect. 

Even when using the best settings, analysis of the collision kernels showed that the ABC flow produced smaller collision kernels for same sized drops. In other DNS studies similar differences for the collision kernels of same sized drops were found. According to \citet{kunnen:13}, this disagreement can be mainly attributed to differences in the radial distribution function among different simulations. Since there is disagreement on monodisperse collision statistics in the literature, it is difficult to make a definite statement about the quality of the ABC results for same sized drops. However, in a comparison of the HR simulations with DNS data it could be seen that the deviations in collision kernel most likely occur only in a narrow size range of about $2 \Delta r$. Since drops with such similar size collide very rarely, it is assumed that this should not affect the simulation significantly in most cases. Furthermore, it is encouraging that the enhancement factor correctly showed an increase in collisions due to turbulence for similar sized drops in line with results from other DNS studies \citep[e.g][]{wang:09}.

The HR runs produced collision kernels very similar to the LR run despite having significantly fewer drops per bin. It is also noteworthy to mention that the 11~realisations of the HR run produced a very small variance in the collision kernel. This suggests that the collision kernel results are statistically robust. However, it also displays some distinct peaks in the enhancement factor for the size ratio of $r_1/r_2=0.93$, which appear in all 11 runs in the same way. This might be an indicator that the trajectories of the drops in the ABC flow were not sufficiently chaotic or stochastic. \citet{wang:91} found that for settling particles, the chaotic behaviour can be reduced in an ABC flow, compared to non-settling particles.

In a comparison of the enhancement factor it could be seen that the ABC flow simulation underestimates the contribution of the turbulent flow compared to the DNS for drop size ratios smaller than~0.6. For the HR runs the average enhancement factor in the interval up to $r_1/r_2=0.6$ was 0.98 with values scattered slightly below and above~1. At the same time the results from the DNS indicated a slight increase in the collision kernel due to turbulence in that size ratio interval, with an average enhancement factor of~1.08. This effect could not be replicated in the same way in the ABC simulation for this size ratio range. The reason for that is that the ABC flow, being stationary and coherent, is more effective in separating drops of different sizes than the DNS flow and therefore reduces their ability to collide. Even though the physical effect is on average opposite in both simulations, the effect on the collision kernel is very small, since both enhancement factors are close to 1 for size ratios below 0.6.

When investigating the contribution of the ABC flow to the collision kernel, the HR runs show a spread in the enhancement factor for drop size ratios larger than 0.7. This spread was caused by an enhancement of collisions for drops of around 32~$\muup$m. It is assumed that the inertial response time of drops of this size favours them in being swept out of the vortices and accumulate in regions of smaller vorticity. They therefore experience increased numbers of collisions due to the larger local drop density, as reported in previous studies \citep[e.g.][]{kunnen:13}.

In general it can be seen that the influence of the turbulence on the collision kernel is independent of drop size and size ratios until a size ratio of around 0.7. For larger size ratios a spread occurs with drops between 30 and 40~$\muup$m most affected.

Despite being stationary and consisting of only vortex tubes with the same length scales, the ABC flow produces very similar results to the DNS simulations by \citet{ayala:08}. Given the much simpler flow it is no surprise that it could not match the DNS results perfectly, but the differences were very small indeed. The structure of the collision kernel was well reproduced for the right set of ABC flow parameters. Those parameters will most likely vary depending on the scenario in question. The relationship between the ABC parameters and the turbulent flow it attempts to imitate are not known a priori. They can be adjusted according to collision kernels known from literature or newly conducted DNS. Our results suggest that parameters close to the Taylor length scale and the RMS velocity of the turbulent flow give good results for drop collision statistics.

If a larger range of sizes needs to be simulated, a superposition of several ABC flows could be tried. Otherwise the one wavelength of a single ABC flow might only properly interact with a specific range of drop sizes and the synergistic effect of a range of turbulent scales would not be reflected in the drop motion.

The main advantage of the ABC flow is that it needs much less computational power than a DNS, freeing up computational resources for other complex processes. E.g., using the ABC flow would allow to simulate larger numbers of drops, larger domain sizes, better cloud physics like thermodynamic effects \citep[e.g.][]{sardina:18}, aerodynamic interactions between drops \citep{wang:05} or the effect of electric charge \citep{harrison:08}.

Implementing other physical processes, like aerodynamic interactions, would be a good test for the ABC flow to investigate how well it can reproduce results of even more complex models.

\section*{Acknowledgments}
This work was supported by the National Center of Meteorology, Abu Dhabi, UAE, under the UAE Research Program for Rain Enhancement Science.

\bibliography{references}

\end{document}